\documentclass[12pt,preprint]{aastex}

\usepackage{epsf}
\usepackage{graphics}
\usepackage{amsmath}

\begin{document}

\newenvironment{inlinefigure}{
\def\@captype{figure}
\noindent\begin{minipage}{0.999\linewidth}\begin{center}}
{\end{center}\end{minipage}\smallskip}

\newenvironment{inlinetable}{
\def\@captype{table}
\noindent\begin{minipage}{0.999\linewidth}\begin{center}}
{\end{center}\end{minipage}\smallskip}

\medskip

\slugcomment{submitted to ApJ}

\shorttitle{}

\received{}

\title{Extrinsic Radio Variability of JVAS/CLASS Gravitational Lenses}

\author{L.V.E. Koopmans} 
\affil{Space Telescope Science Institute, 3700 San Martin Drive, Baltimore, MD 21218\\
California Institute of Technology, 
Theoretical Astrophysics, 130-33, Pasadena, CA 91125\\
Jodrell Bank Observatory, Macclesfield, Cheshire, SK11 9DL, UK\\}
\email{koopmans@stsci.edu}

\medskip

\author{A. Biggs} 
\affil{Joint Institute for VLBI in Europe,
P.O. Box 2, 7990 AA, Dwingeloo, The Netherlands\\
Jodrell Bank Observatory, Macclesfield, Cheshire, SK11 9DL, UK}

\medskip

\author{R.D. Blandford}
\affil{California Institute of Technology, 
Theoretical Astrophysics, 130-33, Pasadena, CA 91125} 

\medskip

\author{I.W.A. Browne, N.J. Jackson, S. Mao, P.N. Wilkinson} 
\affil{Jodrell Bank Observatory, Macclesfield, Cheshire, SK11 9DL, UK}

\medskip

\author{A.G. de Bruyn} 
\affil{ASTRON, P.O. Box 2, 7990 AA, Dwingeloo, The Netherlands\\
Kapteyn Astronomical Institute, P.O. Box 800, 9700 AV Groningen, The Netherlands}

\medskip

\author{J. Wambsganss} 
\affil{Universit\"{a}t Potsdam, Institut f\"{u}r Physik, Am Neuen Palais 10,  
14469 Potsdam, Germany}

\medskip

\begin{abstract}
We present flux-ratio curves of the {\sl fold and cusp} (i.e. close
multiple) images of six JVAS/CLASS gravitational lens systems.  The
data were obtained over a period of 8.5 months in 2001 with the
Multi-Element Radio-Linked Interferometer Network (MERLIN) at 5--GHz
with 50~mas resolution, as part of a MERLIN Key-Project. Even though
the time delays between the fold and cusp images are small ($\la$1\,d)
compared to the time-scale of intrinsic source variability, all six
lens systems show evidence that suggests the presence of extrinsic
variability. In particular, the cusp images of B2045+265 -- regarded
as the strongest case of the violation of the cusp relation (i.e. the
sum of the magnifications of the three cusp images add to zero) --
show extrinsic variations in their flux-ratios up to $\sim$40\%
peak-to-peak on time scales of several months. Its low Galactic
latitude of $b\approx -10^\circ$ and a line-of-sight toward the Cygnus
superbubble region suggest that Galactic scintillation is the most
likely cause. The cusp images of B1422+231 at $b\approx +69^\circ$ do
not show strong extrinsic variability. Galactic scintillation can
therefore cause significant scatter in the cusp and fold relations of
some radio lens systems (up to 10\% rms), even though these relations
remain violated when averaged over a $\la$\,1 year time baseline.
\end{abstract}

\keywords{gravitational lensing --- scattering --- ISM: general }

\section{Introduction}

Cosmological Cold--Dark--Matter (CDM) simulations predict the
existence of condensed structures in the halos around massive galaxies
(e.g. Klypin et al. 1999; Moore et al. 1999), if the initial
power-spectrum does not cut off at small scales and dark matter is
cold and not self-interacting. However, observationally, we see at
most the high-mass tail of these structures in the form of dwarf
galaxies. This raises the question of where most of their less massive
($10^{6}-10^9$~M$_\odot$) counterparts are located. Either (i) these
CDM structures have not formed -- in conflict with CDM predictions --
or (ii) they consist predominantly of dark-matter and baryons have
been blown out, preventing star formation altogether, or (iii) baryons
are present but have not condensed inside their potential well to form
visible stars. If either one of the latter two is the case, the only
way to detected them is through their gravitational effect, in
particular through dynamics and lensing.

The initial suggestion by Mao \& Schneider (1998) that anomalous
flux-ratios in the lens system B1422+231 can be caused by small-scale
mass substructure in the lens galaxy, was recently extended to a
larger -- although still limited -- sample of gravitational-lens
systems with fold and cusp images (Metcalf \& Madau 2001; Keeton 2001;
Chiba 2002; Metcalf \& Zhao 2002; Dalal \& Kochanek 2002; Brada\v{c}
et al. 2002; Keeton et al. 2002). In particular, analyses have focused
on the so called normalized ``cusp relation'', which says that $R_{\rm
cusp}\equiv \Sigma \mu_i/\Sigma |\mu_i|\rightarrow 0$, for the
magnifications $\mu_i$ of the three merging images of a source well
inside the cusp (Blandford 1990; Schneider \& Weiss 1992). A similar
relation holds for the two fold images.  These relations are only two
of many (in fact $\infty$) scaling laws (Blandford 1990). Because
globular clusters and dwarf-galaxies are too few in number to explain
the rate of anomalous flux-ratios and cusp relations, this could be
used as an argument in favor of CDM substructure as the dominant cause
of these apparent anomalies (e.g. Kochanek \& Dalal 2003).

If the observed violations of the cusp relation (i.e. $R_{\rm cusp}\ne
0$), as discussed above, are due to substructure on mass scales of
$10^6$ to $10^9$ M$_\odot$, the effect should be the same for radio
and optical flux ratios (if the latter are available), and it should
be constant in time. However, another possible explanation is
microlensing of stellar mass objects in combination with a smoothly
distributed (dark) matter component (Schechter \& Wambsganss
2002). This one does not require the optical and radio flux ratios to
behave in the same way, and in particular it predicts the optical flux
ratios to {\em change} over time scales of years. Finally, there is
also the possibility that the flux-density and surface brightness
distribution of lensed radio images are affected by the ionized ISM in
the lens galaxy and/or our Galaxy, also leading to changes in the
apparent value of the cusp relation.

Hence, before one can confidently accept the detection CDM
substructure, it requires rigorous testing to see whether observed
flux ratios correspond to magnification ratios or whether they can be
affected by propagation effects (or microlensing). Here, we make the
first coordinated attempt to test the effects of propagation on the
observed {\sl radio} fluxes of lensed images.

In Sections 2 and 3 we present the first results of our MERLIN
Key-Project (Biggs et al. 2003) to search for extrinsic variability
between fold and cusp images (i.e. close multiple images), based on
their flux-density curves. A discussion and conclusions are given in
Sect.4.

\section{MERLIN 5--GHz Data}

MERLIN 5--GHz data were obtained between Feb 21 and Nov 7 2001. A
total of 41 epochs of 24 hours each were obtained, on average once per
week. Eight lens systems were observed (i.e. Table~1 plus B1608+656
and B1600+434) of which seven are four-image systems and one is a
double. The data acquisition and reduction is described in Biggs et
al. (2003), which also presents the flux-density curves of all the
lensed images.

In this paper, we focus on the {\sl flux-ratio} curves. This approach
has several advantages when looking for extrinsic variability. The
dominant errors on flux-density curves in the radio are those
resulting from residual noise in the maps and from multiplicative
errors as a result of erroneous flux calibration. Because
multiplicative errors are equal for each of the lensed images, they
disappear in the flux-ratio curves (not corrected for the time-delays)
which should therefore be flat and dominated by noise in the absence
of variability.

All presented lens systems also have small time-delays between
cusp/fold images ($\la$1\,d) compared to the time between observations
and the time-scale of intrinsic variability as seen in the
flux-density curves (Biggs et al. 2003). Hence, intrinsic flux-density
variations should effectively occur simultaneously in fold and cusp
images and thus disappear in the flux-ratio curves. Throughout this
paper we therefore assume that (i) due to the small time-delays
between fold/cusp images, intrinsic variability does not affect the
flux-ratio curves, (ii) systematic flux-density errors are
multiplicative and also do not affect the flux-ratio curves, and (iii)
extrinsic variability does not correlate between lensed images.

We exclude the double B1600+434 and the quad B1608+656 from our
analysis, which both have non-negligible time-delays (i.e. several
weeks to months; see Fassnacht et al. 1999a, 2002; Koopmans et
al. 2000; Burud et al. 2000).

\section{Results}

\subsection{Normalized Flux-ratio Curves}

\begin{figure*}
\begin{center}
\leavevmode
\hbox{%
\epsfysize=0.8\vsize
\epsffile{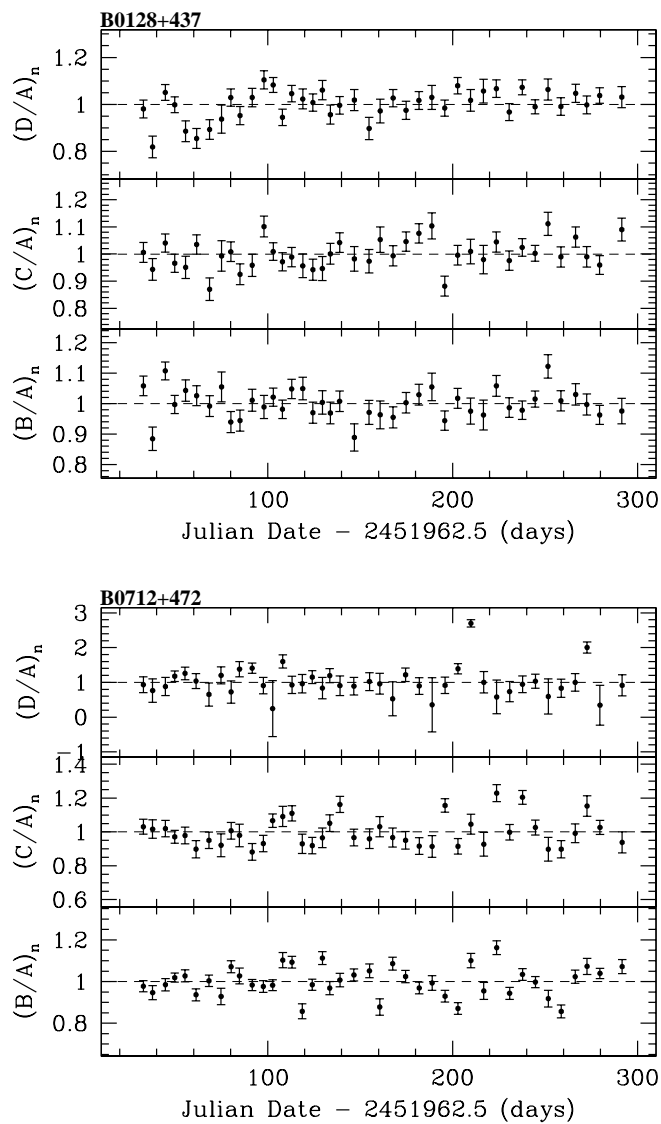}}
\end{center}
\caption{The normalized flux-ratio curves of the three independent
image pairs for all six quadruple lens systems.  The scale on the
y-axis is set to $\pm$5 times the rms scatter of the flux-ratio
curves. The errors on the flux-ratio curves are determined from the
errors on the individual flux-density curves.\label{fig:rcurves}}
\end{figure*}

In Fig.\ref{fig:rcurves} the resulting flux-ratio curves of all images
are shown with respect to image A which is often the brightest
image. We follow the labeling of these images as published in the
literature (e.g. Biggs et al. 2003). Each flux-ratio curve has been
normalized to unity by dividing them through the average flux-ratio of
all 41 epochs. The errors are the square root of the sum of the two
fractional (noise) errors on the flux-densities squared. The
flux-density errors are determined from the rms in the residual maps
(i.e. the radio maps with the lensed images subtracted).

In Table~1, we list (i) the average flux-ratios and the rms scatter
for each image pair, (ii) the reduced--$\chi^2$ values, by assuming
that each normalized flux-ratio should be unity in the case of no
extrinsic variability and under the assumptions mentioned in
Sect.2, and (iii) the values of $R_{\rm cusp}$ (see Mao \& Schneider
1998; Keeton et al. 2002), which we discuss further in Sect.4.

\subsection{Evidence for Extrinsic Variability}

To test for extrinsic variability in the lensed images, on time-scales
less than the monitoring period of 8.5 month, we introduce the
following method:

Let us designate the normalized light curves of the individual
cusp/fold images as $a_n \equiv A/\langle A\rangle$, $b_n \equiv
B/\langle B\rangle$ and $c_n \equiv C/\langle C\rangle$, where their
average flux-densities over the 41 epochs are $\langle A\rangle$,
$\langle B\rangle$ and $\langle C\rangle$, respectively\footnote{We
use the notation $A$, $B$, $C$ and $a_n$, $b_n$, $c_n$ to indicate
both the light curves as a whole, as well as their individual
flux-density values.}

First, the points $(a_n,b_n,c_n)$ are plotted in a three-dimensional
Cartesian space, such that multiplicative errors {\sl and} intrinsic
flux-density variations (the latter because of the negligible
time-delays) move points parallel to the vector $(1,1,1)$.

Second, each point $(a_n,b_n,c_n)$ is projected on to a
two-dimensional plane that is normal to the vector $(1,1,1)$. Hence,
the projected points will {\sl not} move on that plane, either because
of intrinsic flux-density variations or multiplicative errors. Both of
these are movements perpendicular to the plane and thus translate to
the same projected point.

Third, if one defines the $x$--axis, $\hat x$, of this two-dimensional
plane to be the projected $a$--axis, $\hat a$, of the
three-dimensional space, and $\hat y$ to be perpendicular to $\hat x$ in the
same normal plane, one finds the following simple mapping:
\begin{eqnarray}
     x &=& (2 a_n - b_n -c_n)/\sqrt{6} \nonumber\\
     y &=& (b_n - c_n)/\sqrt{2}
\end{eqnarray}
or in polar coordinates
\begin{eqnarray}
  r^2      &=& {x^2 + y^2}\nonumber\\
  \theta &=& \arctan(x,y).
\end{eqnarray}
Because $\hat a$ projects on to $\hat x$, any uncorrelated
extrinsic variations in image A will only result in a movement of a
point along $\hat a$ and thus only along the $\hat x$ axis.

Because the 1--$\sigma$ errors on the normalized flux-densities $a_n$,
$b_n$ and $c_n$ are known from the observations, one can calculate the
corresponding expected 1--$\sigma$ errors on $x$ and $y$.
\begin{eqnarray}
    \sigma_x^2 & = & {2/3\, \sigma_{a}^2 + 1/6\, \sigma_{b}^2 + 1/6\, \sigma_{c}^2}\nonumber\\
    \sigma_y^2 & = & {1/2\, \sigma_{b}^2 + 1/2\, \sigma_{c}^2}
\end{eqnarray}
and similarly
\begin{eqnarray}
    \sigma_r^2 & = & ((2 a_n - b_n -c_n)^2\, \sigma_a^2 + (2 b_n - c_n -a_n)^2\, 
                    \sigma_b^2 + (2 c_n - a_n -b_n)^2\, \sigma_c^2)/(9 r^2).
\end{eqnarray}
Notice that the scatter in $x$ will be a combination of the scatter in
$a_n$, $b_n$ and $c_n$, if each image behaves independently.

On the other hand,
\begin{equation}
  \chi^2_r = \frac{1}{\rm DOF}\sum_i{(r_i/\sigma_{r,i})^2}
\end{equation}
is a direct estimator of the significance of the presence of extrinsic
variability on time-scales of $<$8.5 month, {\sl irrespective of the
image(s) it occurs in}. In other words, it does not tell us which
image or images exhibit extrinsic variability, only that extrinsic
variability is present if $\chi^2_r$ is significantly larger than
unity.

The significance of extrinsic variability in individual image is far
more difficult to assess.  However, we can estimate the level of
extrinsic variability in image A, for example, by knowing that the 
expected variance in that image, due to noise {\sl and} in the absence of
extrinsic variability, should be
\begin{equation}
  {\rm E}\{\langle \sigma_a^2 \rangle\} \approx 3/2\, {\rm var}(x) - 1/2\, {\rm var}(y).
\end{equation}
If the observed value of $\langle \sigma_a^2 \rangle = (\sum^N_i \sigma_{a,i}^2)/{N}$ is smaller than 
${\rm E}\{\langle \sigma_a^2 \rangle\}$, the difference is due to extrinsic
variability with an estimated  variance of
\begin{equation}
  {\rm var}(a_{\rm ext}) \approx {\rm E}\{\langle \sigma_a^2 \rangle\} - \langle \sigma_a^2 \rangle.
\end{equation}
The same procedure can be repeated for each of the other images.  In
Table~2, we have listed the values of $\chi^2_r$ and the values of
${\rm var}(a_{\rm ext},b_{\rm ext},c_{\rm ext})$, if larger than zero
(note that ${\rm E}\{\langle \sigma_a^2 \rangle\}$ is an estimate and
could therefore be smaller than $\langle \sigma_a^2 \rangle$ when
measured from a finite set of observations).

Finally, we further discuss whether correlations between the flux
measurements of the merging images could potentially occur. We note,
however, that $\sigma_{a/b/c}$ are noise errors as determined from
residual maps, i.e. the original maps after we subtract of the
best-fit model of the lensed images. The residual radio maps are
consistent with noise maps. Since the images are separated by many
beam sizes (i.e. resolution elements), the flux measurements of images
A, B and C -- even though measured from the same map -- are
independent, except for the multiplicative errors as explained
previously. Hence there should be {\sl no} effect of measurement
correlations in Eq.(3) or (4), which could skew our results.

Hence, the technique discussed above is explicitly designed to
separate the effects of multiplicative errors, extrinsic variability
and noise, and should also be free of measurement correlations. For
example, if one were to cross correlate (e.g.~using the Spearmann rank
correlation) the flux-ratio curves (Fig.1) of a single lens system
with each other, one would find that they correlate strongly, even in
the absence of extrinsic variability. The reason being that the same
noise variations in image A would be introduced in both $(B/A)_{\rm
n}$ and $(C/A)_{\rm n}$. A Spearman rank correlation on flux-ratio
curves without extrinsic variability but with similar noise properties
and number of epochs confirms this. However, one notices from
equations 3 and 4 that any multiplicative error does not affect
$\sigma^2_{x/y}$ or $\sigma^2_r$ (where it cancels out) or the
projection on the plane that we defined in equations 1 and 2, as
previously discussed. In addition, one finds from equations 1, 2 and 4
that if there is no extrinsic variability, $\chi^2_r \rightarrow 1$,
whereas the presence of extrinsic variability implies $\chi^2_r > 1$.
Hence, $\chi^2_r$ is indeed independent from multiplicative errors and
therefore the correct estimator of the significance of the presence of
extrinsic variability, in the (shown) absence of measurement
correlations.

\subsection{Individual Lens Systems}

Here, we discuss each case, based on their reduced $\chi^2$
values. Image D is not considered because of its faintness and larger
inferred time-delay compared with the other images.

\subsubsection{All systems, except B2045+265} 

Based on the relatively low values of $\chi^2_r$ and the estimated
levels of extrinsic variability (Table~2) for the images of B0128+437,
B1359+154 and B1422+231 and B1555+375, and the remaining possibility
that some minor undetected additive errors could be present, the
evidence for extrinsic variability in these four systems is not
totally convincing. We exclude these from further discussion.

In the case of B0712+472, the reduced $\chi^2$ values of the
$(B/A)_{\rm n}$ flux-ratio curves and also $\chi_r^2$ seem more 
significant. In Fig.1 we see that a large number of epochs are deviant
over the entire observing season. Deviations of the $(C/A)_{\rm n}$
flux-ratio curve from unity are less significant, probably because
image C has a larger fractional error than images A and B. Even though
there is some evidence in this system for extrinsic variability
between the two fold images, we conservatively regard it also as weak
and we will concentrate our discussion on B2045+265. In Sect.4,
however, we further discuss what a possible reason for some of the
higher values of $\chi_r^2$ and extrinsic variability can be.

\subsubsection{B2045+265} 

In Tables 1 and 2, we see that (i) both the $(B/A)_{\rm n}$ and
$(C/A)_{\rm n}$ flux-ratio curves have very high values of the reduced
$\chi^2$, reflected also in large rms values, (ii) the estimated rms
values of extrinsic variability and the value of $\chi_r^2$ are
very large, and (iii) a visual inspection of the $(B/A)_{\rm n}$ and
$(C/A)_{\rm n}$ flux-ratio curves shows changes of up to $\sim$40\% on
time scales of several months. Because the time delays between the
cusp images are only a fraction of a day (Fassnacht et al. 1999b),
residual intrinsic source variability can not cause these variations.

A more quantitative analysis based on the structure function
(e.g. Simonetti et al. 1985) of the flux-ratio curves (indicated by
$R(t)$) is shown in Fig.2. The structure function
$<D^{(1)}(\tau)>=<[R(t+\tau)-R(t)]^2>$ quantifies the average rms
fluctuations (squared) between two points on the same flux-ratio
curve, separated by a time lag $\tau$. A lower value of
$<D^{(1)}(\tau)>$ means a stronger correlation (assuming no errors).
Fig.2 shows that, even though $<D^{(1)}(\tau)>$ fluctuates
considerably, it continues to increase toward longer lags. Around
$\tau\sim 150$~d, the rms suddenly decreases considerably, suggesting
possible long-term correlated variations in the flux-ratios on that
time-scale. If $<D^{(1)}(\tau)>$ increases beyond $\tau\ga 200$~days,
flux-density variations of several tens of percent on a time-scale of
$\ga 1$\,yr could be present as well. However, we note that the
overlap of the flux-ratio curves becomes smaller for longer lags and
consequently the errors become larger. Longer observations are
required to make stronger statements about the longer time lags. Even
so, similar fluctuations of the structure function are seen in other
scintillating sources (e.g. Dennett-Thorpe \& de Bruyn 2003).

Several reliability checks of the extrinsic variations of the cusp
images of B2045+265 are called for: First we note that the lensed
images are of roughly equal brightness and -- within a factor about
two -- as bright as the images in B0128+437, B1359+154, B0712+472 and
B1555+375. Hence there is no indication that the observed flux-ratio
variations are related to the faintness or brightness of the lensed
images. Second, there are no problems with the closeness between the
cusp images ($\ga 0.3''$) and the separation of their flux densities
because of the high resolution of the MERLIN radio maps
($\sim$50~mas). Hence the fluxes of the three images are fully
independent. Third, we have calculated the Spearman rank correlation
coefficients ($r_{\rm S}$) between each of its images A, B and C and
those of the other five lens systems. This leads to 45 independent
values of $r_{\rm S}$ (i.e. noise does not introduce correlation in
this case), which on average should tend to zero. We find $<r_{\rm
S}>$=0.0024 and an rms of 0.153. The theoretical expectation value of
the rms value is $1/\sqrt{N-1}$=0.151, where $N=45$ in our
case. Hence, we recover the expectation values of both the average and
rms. This shows that any correlation between the images can not be the
result of obvious systematic errors in the data-reduction process, in
the creation of the flux-ratio curves, or in our analysis.

Hence, we confidently conclude that the cusp images of B2045+265 show
strong evidence for the presence of extrinsic variability.

\section{Discussion \& Conclusions}

We have presented the flux-ratio curves of six gravitational lens
systems, each composed of 41 epochs taken over a period of 8.5 months
in 2001 with MERLIN at 5 GHz, as part of a MERLIN Key-Project.  The
systems were chosen to have merging cusp or fold images, such that the
time-delays between these images are negligible ($\la$1\,d) compared
to the time-scale of intrinsic variability and the rate at which the
light curves are sampled. The flux-ratio curves should therefore be
void of intrinsic variability and multiplicative errors. The main goal
of our program was to find additional cases of extrinsic variability
other than radio-microlensing in B1600+434 (Koopmans \& de Bruyn 2000,
2003).

We find some statistical evidence for extrinsic variability in all six
lens systems, based on reduced $\chi_r^2$ values larger than unity
(Sect.3.2; Tables~1 \& 2). Residual intrinsic variations -- due to the
finite time delays -- or small additive error are unlikely to be the
cause of this, but can not fully be excluded yet. The high resolution
of MERLIN also ensures negligible correlations between the fluxes of
the merging images. The evidence for B0128+437, B1359+154, B1422+231
and B1555+375 is fairly marginal. The case for B0712 is stronger,
however, and this object clearly deserves further study. The best case
is B2045+265, which we discuss further below.

Even though radio-microlensing cannot be excluded, we think at this
point that Galactic scintillation is the more likely cause of some of
the higher values of $\chi^2$ (Tables~1 \& 2). Indeed, {\sl all} compact
extragalactic radio sources should show refractive scintillation at
some level. At wavelengths of 5 GHz and for image sizes $\sim$1 mas,
the expected rms fluctuations due to scintillation, in a typical
line-of-sight out of the Galactic plane, are a few percent
(e.g. Walker 1998, 2001), which are comparable to the observed
flux-density errors.

One gravitational lens systems, B2045+265, shows unambiguous evidence
for extrinsic variability, based on the reduced $\chi^2$ values
significantly larger than unity (Tables~1 \& 2) and visually apparent
long-term variations in its flux-ratio curves (Fig.1). One possible
explanation for the variations is radio microlensing similar to
B1600+434 (Koopmans \& de Bruyn 2000, 2003). However, because
B2045+265 has a Galactic latitude $b \approx -10^\circ$ and is the
lowest Galactic latitude system in our sample, Galactic refractive
scintillation is the more likely explanation.

To examine this, first we naively use the revised electron-density
model of our Galaxy by Cordes \& Lazio (2003). This model gives a
scattering measure of $8\times 10^{-4}$~kpc~m$^{-20/3}$, an angular
broadening at 5~GHz of 50~$\mu$as and a transition frequency of 22~GHz
between the weak and strong scattering regimes. If we choose the
source size to be 250~$\mu$as, we find a modulation index of 7\% (from
Walker 1998, 2001) or an rms scatter of $\sim$10\% in the flux-ratio
curves (as observed; Table~1), and a typical variability time-scale of
$\sim$1~week for an effective transverse velocity (i.e. the velocity
of the ISM, earth, local and solar peculiar motions combined) of the
medium of 50~km\,s$^{-1}$. Note however that the time-scale of
variability might vary with time of the year due to the earth's motion
(e.g. Dennett-Thorpe \& de Bruyn 2000, 2002).

Refractive scintillation could therefore explain the observed
extrinsic variations up to a time-scale of possibly several weeks in
B2045+265 for reasonable lensed-images sizes. However, the structure
function shows correlated variations on time-scales that are much
longer. These could either indicate modification(s) of the Kolmogorov
spectrum of density fluctuations that was assumed in the above
calculation or a very low transverse velocity of the medium,
i.e. 10~km\,s$^{-1}$. If there is more power in the spectrum on larger
scales, or a cutoff on smaller scales, fluctuations will become
stronger on longer time scales (e.g. Blandford et al. 1986; Romani,
Narayan \& Blandford 1986; Goodman et al. 1987). Such large-scale
electron density waves might also explain the apparent fluctuations in
the observed structure function (Fig.2).

On further examination, however, we find that B2045+265 is very close,
if not seen through, the Cygnus ``superbubble'' region (see Fig.6 in
Fey, Spangler \& Mutel 1989), making our analysis based on the model
in Cordes \& Lazio (2003) rather uncertain. This region has
considerably enhanced scattering measures and if this is the case for
B2045+265 as well, it would strongly support Galactic scintillation as
the cause of the observed flux-density variations. The complexity of
such regions, where turbulence in the ionized ISM presumably
originates, could be the reason why we see large-amplitude
fluctuations in the flux ratios with time scales that are not expected
from simple Kolmogorov turbulence models (see also e.g. J1819+3845;
Dennett-Thorpe \& de Bruyn 2000, 2002).

Finally, it is interesting to note that B2045+265 has the strongest
and most significant violation of the cusp relation of all
known lens systems (Keeton et al. 2003). Even so, the values of
$R_{\rm cusp}$ of the systems discussed in this paper (see Table~1)
agree with those in Keeton et al. (2003). However, the strong observed
variations in the flux-ratio curves should caution against the
use of both flux-ratios and values of $R_{\rm cusp}$ (Sect.1) --
derived from single-epoch observations -- even if the inferred
time-delays are only a few hours! 

Whether the violation of the cusp relation in B2045+265, averaged over
8.5 months (Table~1), and Galactic refractive scintillation and/or
scattering is completely coincidental, is not clear at this point. At
any instant in time, however, large scale electron-density
fluctuations in the Galactic ISM can focus {\sl or} defocus the
images with long time scales of variability -- as is apparent from our
observations -- probably even more so toward regions of enhanced
turbulence (i.e. the Cygnus region). CDM substructure mostly focuses
the images. It is interesting to note that B0712+472, probably the
system with second-best evidence for extrinsic variability in our
sample, also has a low Galactic latitude $b=+23^\circ$.

While the observations reported in this paper do not contradict the
exciting conclusion that CDM substructure might have been detected
within the central regions of lens galaxies, they do suggest that
extrinsic, refractive effects are also of importance and that it is
imperative to carry out further, multi-frequency monitoring to
distinguish them from achromatic, gravitational effects.

{\acknowledgments LVEK acknowledges the support from an STScI
Fellowship grant. RDB is supported by an NSF grant
AST--0206286. MERLIN is a National Facility operated by the University
of Manchester at Jodrell Bank Observatory on behalf of PPARC.  LVEK
thanks Chris Kochanek and Neal Dalal for suggestions that improved the
presentation of this work. We thank the referee for helping to clarify
the manuscript.}

\clearpage

\begin{table*}
\begin{center}
\begin{tabular}{cccccc}
\hline
\hline
    & $<r_{(B/A)}>$ & $<r_{(C/A)}>$ & $<r_{(D/A)}>$ & $\chi^2$/DOF & $R_{\rm cusp}$ (ABC) \\
\hline
B0128+437 & 0.584(0.029) & 0.520(0.029) & 0.506(0.032) & 1.8/1.9/2.4 & 0.445 (0.018) \\
B0712+472 & 0.843(0.061) & 0.418(0.037) & 0.082(0.035) & 4.8/3.2/8.0 & 0.255 (0.030) \\
B1359+154 & 0.580(0.039) & 0.782(0.031) & 0.193(0.031) & 1.9/0.9/1.2 & 0.510 (0.024) \\
B1422+231 & 1.062(0.009) & 0.551(0.007) & 0.024(0.006) & 1.8/2.0/1.5  & 0.187 (0.004) \\
B1555+375 & 0.620(0.039) & 0.507(0.030) & 0.086(0.024) & 3.4/2.1/2.4 & 0.417 (0.024) \\
B2045+265 & 0.578(0.059) & 0.739(0.073) & 0.102(0.025) & 8.2/10.9/2.9 & 0.501 (0.035) \\
\hline
\hline
\end{tabular}
\end{center}
\noindent{\flushleft Table~1.--- \small The flux-ratios of each image
pair. The rms scatter in the flux-ratio is indicated between
parentheses, calculated from the 41 epochs. The reduced $\chi^2$
values are listed as well, calculated on the basis that each
normalized flux-ratio curve should be unity and that there is no
variability. In addition, the values of $R_{\rm cusp}$ (see Sect.1)
and its rms (between parentheses) are listed (e.g. Mao \& Schneider
1998; Keeton et al. 2002).}
\end{table*}

\begin{table*}
\begin{center}
\begin{tabular}{ccccc}
\hline
\hline
    & rms($a_{\rm ext}$) & rms($b_{\rm ext}$) & rms($c_{\rm ext}$) & $\chi^2_r/{\rm DOF}$ \\
\hline
B0128+437 & 2.9\% & 1.9\% & 2.5\% & 3.3 \\
B0712+472 & 4.8\% & 4.2\% & 4.8\% & 6.2 \\ 
B1359+154 & 1.0\% & 4.6\% &  --   & 2.8 \\
B1422+231 & --    & 0.6\% & 0.9\% & 3.7 \\
B1555+375 & 3.3\% & 4.2\% & 3.0\% & 5.3 \\
B2045+265 & 6.1\% & 7.0\% & 7.2\% & 17.1\\
\hline
\hline
\end{tabular}
\end{center}
\noindent{ Table~2.--- \small The estimated rms levels of extrinsic
variability in images A, B and C. The reduced values of $\chi^2_r$ are
given to indicate the significance of the presence of extrinsic
variability in the combined set of images. The dashes indicate that
the estimated variance was smaller than zero (see Sect.3.2 for more
details).}
\end{table*}

\clearpage

\clearpage

%\begin{figure*}
\begin{center}
\leavevmode
\hbox{%
\epsfysize=0.8\vsize
\epsffile{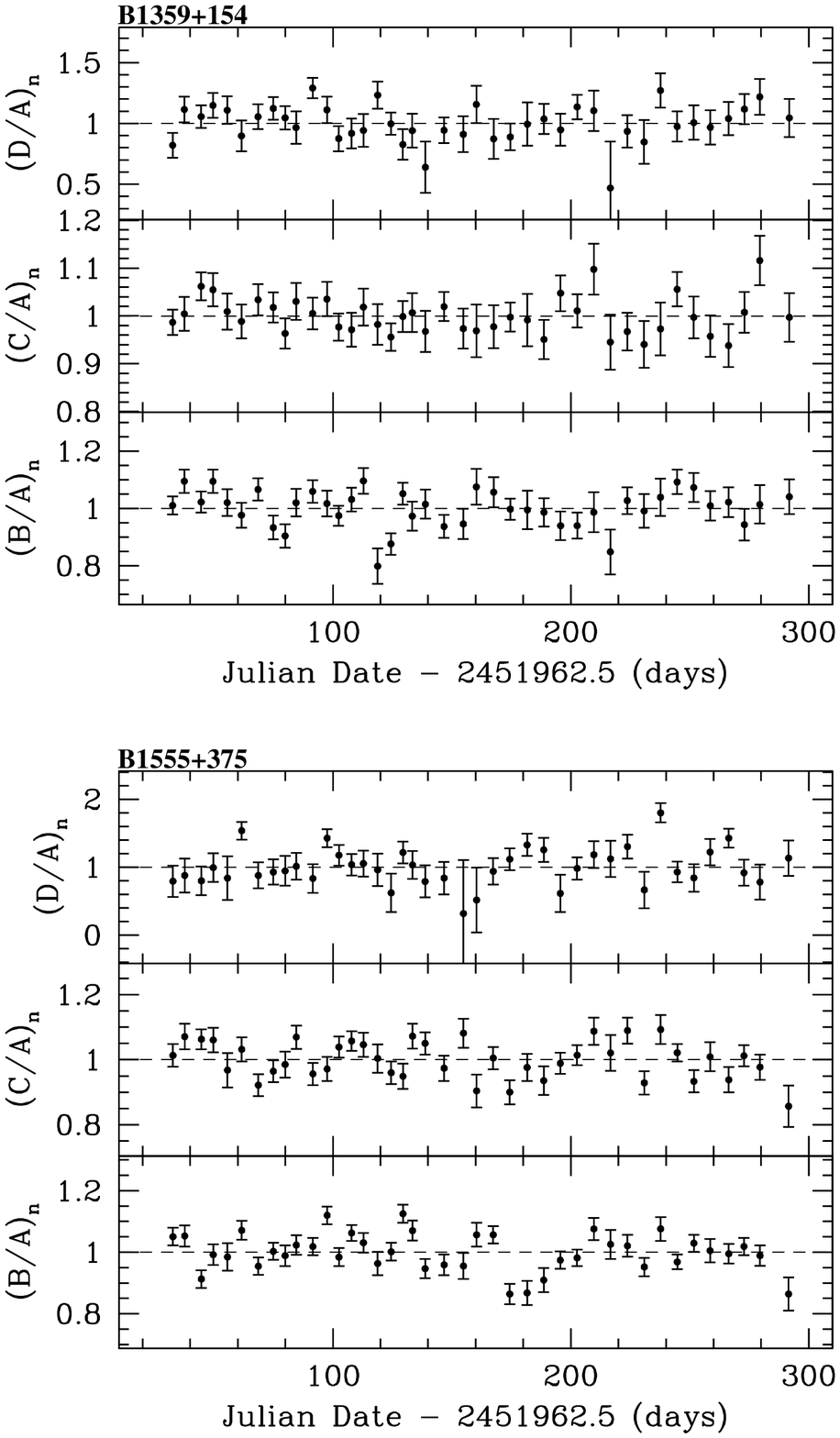}}
\end{center}
%\caption
{Fig. 1.--- Continued}
%\end{figure*}

\clearpage

%\begin{figure*}
\begin{center}
\leavevmode
\hbox{%
\epsfysize=0.8\vsize
\epsffile{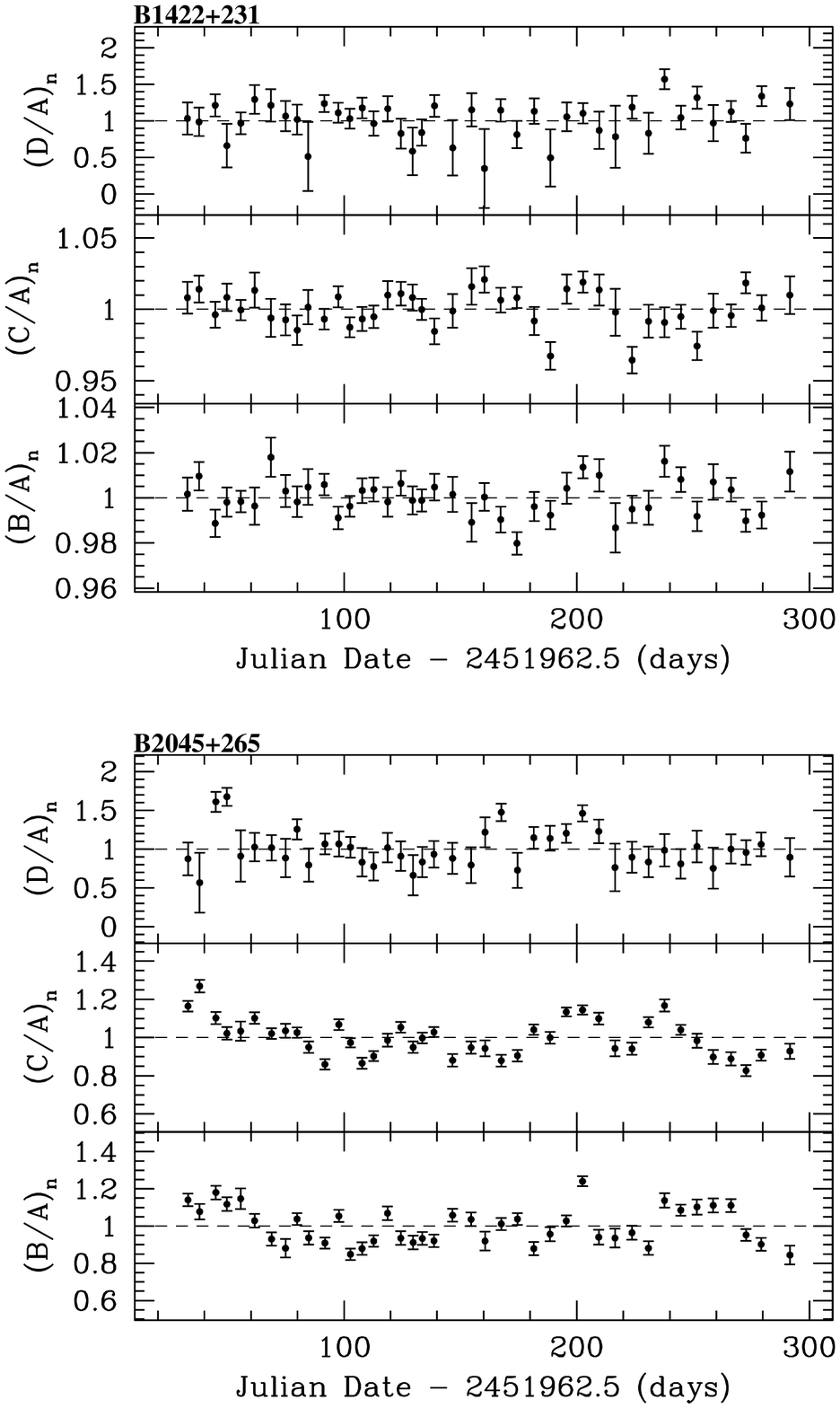}}
\end{center}
%\caption
{Fig. 1.--- Continued}
%\end{figure*}

\clearpage

\begin{inlinefigure}
\begin{center}
\resizebox{\hsize}{!}{\includegraphics{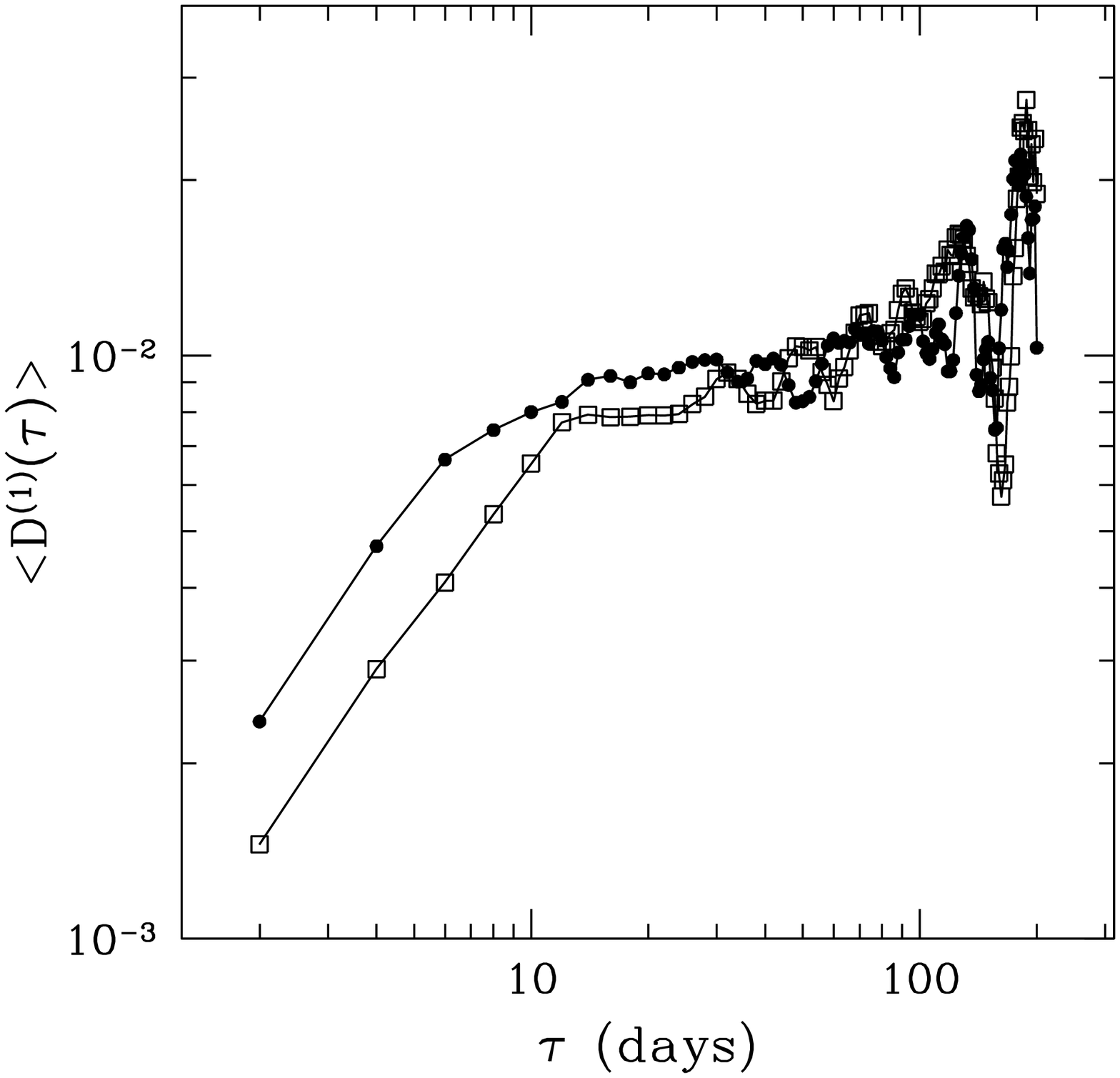}}
\end{center}
\figcaption{The structure functions of the normalized flux-ratio
curves $(B/A)_{\rm n}$ (circles) and $(C/A)_{\rm n}$ (open squares) of
B2045+265 between time-lags of 2 and 200 days. The break below
$\sim$10\,d is the result of variations in the flux-ratio curves that
correlate on that time scale, but are also affected by the on-average
1-week time separation between observations. }
\end{inlinefigure}

\end{document}